\documentclass[]{spie}  
\usepackage[]{graphicx}
\usepackage{amsmath,amssymb,amsfonts}
\usepackage{algorithmic}
\usepackage{graphicx}
\usepackage{textcomp}

\usepackage{hyperref}

\title{Comparison Different Vessel Segmentation Methods in Automated Microaneurysms Detection in Retinal Images using Convolutional
Neural Networks} 

\author{Meysam Tavakoli\supit{a} and Mahdieh Nazar\supit{b}
\skiplinehalf
\supit{a}Department of Physics, Indiana University-Purdue University, Indianapolis, IN, USA 46202; \\
\supit{b}Department of Biomedical Sciences, Shahid Beheshti Medical School, Tehran, IRAN 
}

\begin{document} 
  \maketitle 

To appear in:
Proceedings Volume 11317, Medical Imaging 2020: Biomedical Applications in Molecular, Structural, and Functional Imaging; 113171P (2020) https://doi.org/10.1117/12.2548359
Event: SPIE Medical Imaging, 2020, Houston, Texas, United States

\begin{abstract}
Image processing techniques provide important assistance to physicians and relieve their workload in different tasks. In particular, identifying objects of interest such as lesions and anatomical structures from the image is a challenging and iterative process that can be done by computerized approaches in a successful manner. Microaneurysms (MAs) detection is a crucial step in retinal image analysis algorithms. The goal of MAs detection is to find the progress and at last identification of diabetic retinopathy (DR) in the retinal images. The objective of this study is to apply three retinal vessel segmentation methods, Laplacian-of-Gaussian (LoG), Canny edge detector, and Matched filter to compare results of MAs detection using combination of unsupervised and supervised learning either in the normal images or in the presence of DR. The steps for the algorithm are as following: 1) Preprocessing and Enhancement, 2) vessel segmentation and masking, 3) MAs detection and Localization using combination of Matching based approach and Convolutional
Neural Networks. To evaluate the accuracy of our proposed method, we compared the output of our method with the ground truth  that collected by ophthalmologists. By using the LoG  vessel segmentation, our algorithm found sensitivity of more than 85\% in detection of MAs for 100 color images in a local retinal database and 40 images of a public dataset (DRIVE). For the Canny vessel segmentation, our automated algorithm found sensitivity of more than 80\% in detection of MAs for all 140 images of two databases. And lastly, using Matched filter, our algorithm found sensitivity of more than 87\% in detection of MAs in both local and DRIVE datasets.
\end{abstract}



\section{INTRODUCTION}
\label{sec:intro}  

Diabetic retinopathy (DR) is a microvascular complication of diabetes which is the most usual cause of blindness and vision-loss in the working age people of the western world~\cite{lee2015epidemiology}. 
It has been proved that early detection of DR helps prevent vision lost and blindness~\cite{tavakoli2017automated-onh}. 
DR is a silent disease and may only be recognized by the patient when the changes in the retina have progressed to a level, that treatment is complicated and nearly impossible~\cite{lukac2006color, tavakoli2017comparing}. In general, the severe progression of diabetes is one of the greatest challenges to current health care~\cite{diabetes1993effect}. One of these challenges is the number of people afflicted which continues to grow at an alarming rate~\cite{thomas2015prevalence, pedro2010prevalence}. However, only one half of the patients are aware of the disease.
Establishing of  automated DR detection systems has received a lot of attention from the research community. The computer techniques are applied for providing physicians assistance at any time and to relieve their work load or iterative works as well, to identify object of interest such as lesions and anatomical structures from the image~\cite{tavakoli2017automated-fovea, tavakoli2017automated-onh}. One of the most important steps in the automated screening of DR is the detection of microaneurysms (MAs) ~\cite{tavakoli2013complementary}. MAs are small outpouchings in capillary vessels~\cite{niemeijer2009retinopathy}. They are amongst the first signs of the DR~\cite{abramoff2008evaluation}.
Detection of MAs is the main step in the automated detection of DR screening systems. They are visible immediately after the arterial phase of fluorescein angiography ~\cite{tavakoli2013complementary, walter2007automatic}. Furthermore, counting of MAs has been used as a tool for evaluation of the progression of the DR~\cite{tavakoli2013complementary}.
The objective of this study is to apply three retinal vessel segmentation methods~\cite{tavakoli2017automated-onh, tavakoli2017effect}, 1) Laplacian of Gaussian (LoG) edge detector, 2) Canny edge detector, and 3) Matched filter edge detector to compare detection results of MAs using combination of unsupervised and supervised learning either in normal fundus images or in presence of retinal lesion like in DR. After vessel segmentation and masking using each of these three methods, by combination of Matching based method and Convolutional
Neural Networks (CNNs)~\cite{vedaldi2015matconvnet}  a new hierarchical method is proposed for detection of all MAs. 

There are several studies for the automatic detection of MAs in color retinal images. 
These methods can be generally categorized into three different approaches  including morphological operation, template matching, and supervised learning~\cite{tavakoli2013complementary, niemeijer2005automatic, dashtbozorg2018retinal, habib2017detection, walter2007automatic, quellec2008optimal, wang2016localizing, zhang2010detection, gegundez2017tool, ram2010successive, lazar2012retinal, mizutani2009automated, dai2018clinical}.
Different adjustment was applied based on morphology approach to increase the detection accuracy~\cite{fleming2006automated, mizutani2009automated}. Although this type of processing typically is fast and easy to apply, the ability of the approach is limited by the its builder. In better words, some main hidden structures and uncover patterns could be ignored by the builder and cause false segmentation~\cite{dai2018clinical}. 
Several other mathematical morphology based methods proposed for the detection of red lesions.
According to statistical result, the intensity distribution of MAs is matched to Gaussian distribution~\cite{quellec2008optimal, zhang2010detection, lazar2012retinal}. Therefore, template matching based MA detection approaches were proposed and greatly improved the detection accuracy. 

By growing of machine learning ideas~\cite{tavakoli2019pitching, tavakoli2019bayesian}, studies on MA detection using classification based approaches are mostly seen recently. Antal et al.~\cite{antal2012ensemble} applied a rule-based expert system for MAs detection. In this approach, after selection of MA candidates from retinal images, a rule-based classifier is applied to find true MAs.
Niemeijer et al.~\cite{niemeijer2005automatic} used a hybrid strategy using both top-hat based approach  and a supervised classification. In this approach, MA candidates same as  Antal et al.~\cite{antal2012ensemble} were first selected and then a classifier was trained to differentiate true MAs from false ones. although these machine learning based approaches succeed in detecting hidden structures of features and MAs, they still rely on manually selected features and empirically determined parameters~\cite{dai2018clinical}.
Furthermore, related to third category of detection, learning based approaches,~\cite{gulshan2016development, zhou2017automatic, gargeya2017automated, seoud2015red, abramoff2016improved, orlando2018ensemble, chudzik2018microaneurysm, dai2018clinical, costa2018weakly} proposed to address above issues. Gulshan et al.~\cite{gulshan2016development} presented a deep learning based algorithm to automatically grade DR in retinal images. In this method, deep neural network is applied to process directly the images and output the grading result of DR. While this work successfully addressed the problem of finding hidden structures and empirically determined parameters is not need, for training purpose the classifiers it requires large amounts of retinal images and their annotations, which is costly and time-consuming. Moreover, there is not any  quantitative results generated explicitly for a specific MA. In fact for understanding of the development of DR and monitoring its progress, these quantitative data are critical.
Seoud et al.~\cite{seoud2015red} proposed a novel method for automatic detection of both MAs in color retinal images. The main focus of their work is a new set of shape features, called Dynamic Shape Features, that do not need to precisely segment of red lesion regions. The approach detects all types of MAs  in the images, without distinguishing between them. Differentiating between these lesion types is really important in the clinical practice. The detection of MA is not a practical solution in the medical field. 
Haloi~\cite{haloi2015improved} implemented five layers deep learning with drop out mechanism for diagnosing of early stage DR. The shortcome of the approach was the requirement for a large amount of training data and time-consuming~\cite{wu2017automatic}. 
In comparison, our work focuses on detecting and distinguishing MAs in fundus images. Here, before working on post processing step which we are using the concept of deep learning we add preprocessing unsupervised steps to have some candidates as the MAs and among them using the deep learning we are looking for final true MAs.

\section{Methods} 
\begin{figure}
   \begin{center}
   \begin{tabular}{c}
   \includegraphics[height=10cm, width=9cm]{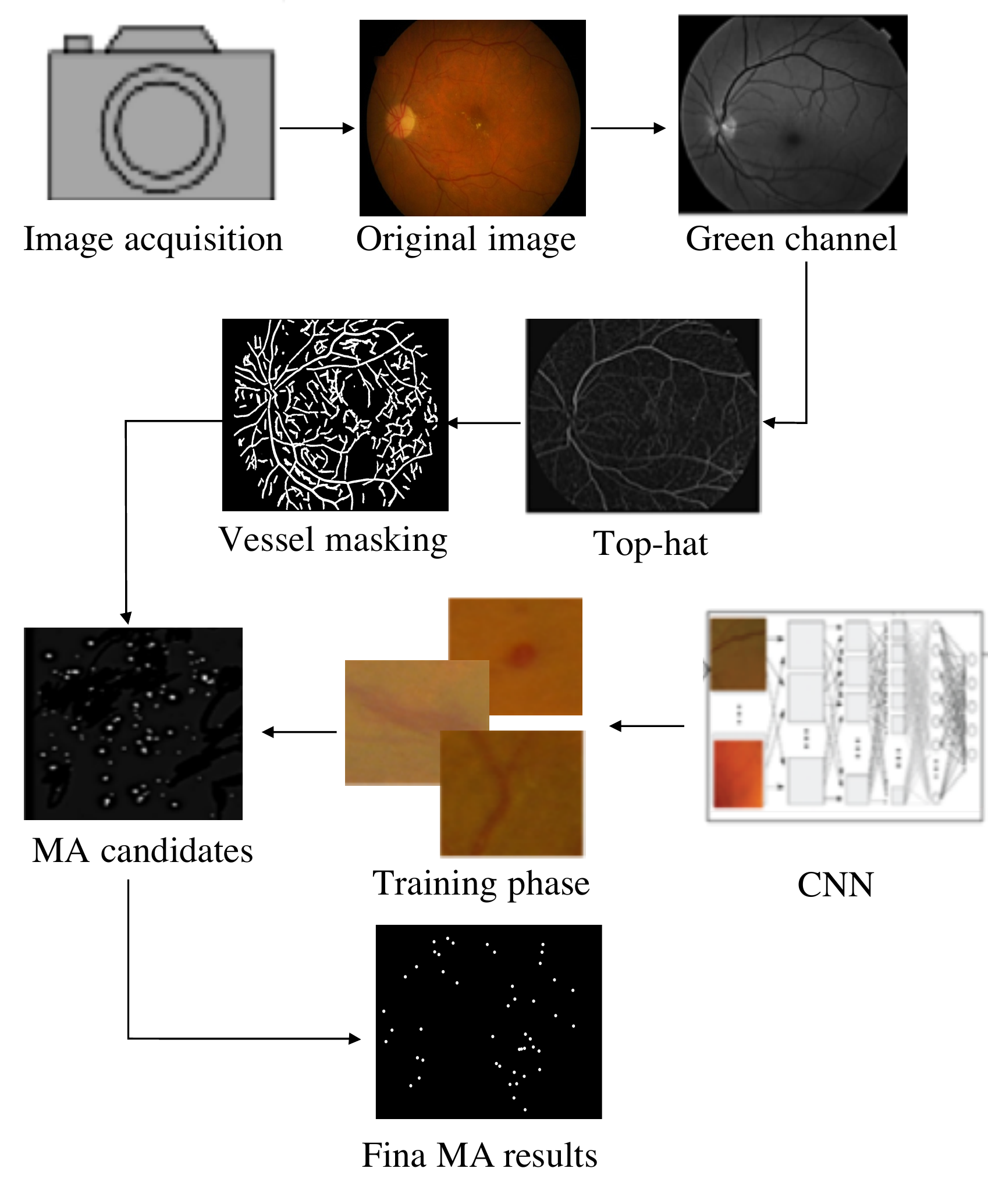}
   \end{tabular}
   \end{center}
   \caption[example] 
   { \label{fig:flowchart} 
Block diagram of proposed method. Here we pictorially show the steps of our algorithm. }
   \end{figure}

A pictorial flowchart of the proposed method is shown in Fig. \ref{fig:flowchart} and the individual processing steps are detailed in the following sub-sections.

\subsection{Materials} 
\label{sec:Materials}

To detect the MAs, two databases (one rural and one publicly available databases) were used. The first rural database was named Mashhad University Medical Science Database (MUMS-DB). The MUMS-DB provided 100 retinal images including 80 images with DR and 20 without DR. The images were obtained 
at 50 degree field of view (FOV) and mostly obtained from the posterior pole view (including ONH and macula) with of resolution $2896 \times 1944$ pixels~\cite{tavakoli2011automated, pourreza2014computationally, tavakoli2011radon}. The second dataset was the DRIVE database consisting of 40 images with image resolution of $768\times 584$ pixels in which 33 cases did not have any sign of DR and 7 ones showed signs of early or mild DR with a 45 degree FOV. For algorithms that operate in a supervised manner this database is often divided into a testing and training set, each containing 20 images. For the test set, two specialists provided manual segmentations for each image. The training set has manual segmentations made by just the first specialist \cite{niemeijer2004comparative}.

\subsection{Preprocessing and Image Enhancement} 
\begin{figure}
   \begin{center}
   \begin{tabular}{c}
   \includegraphics[height=6cm]{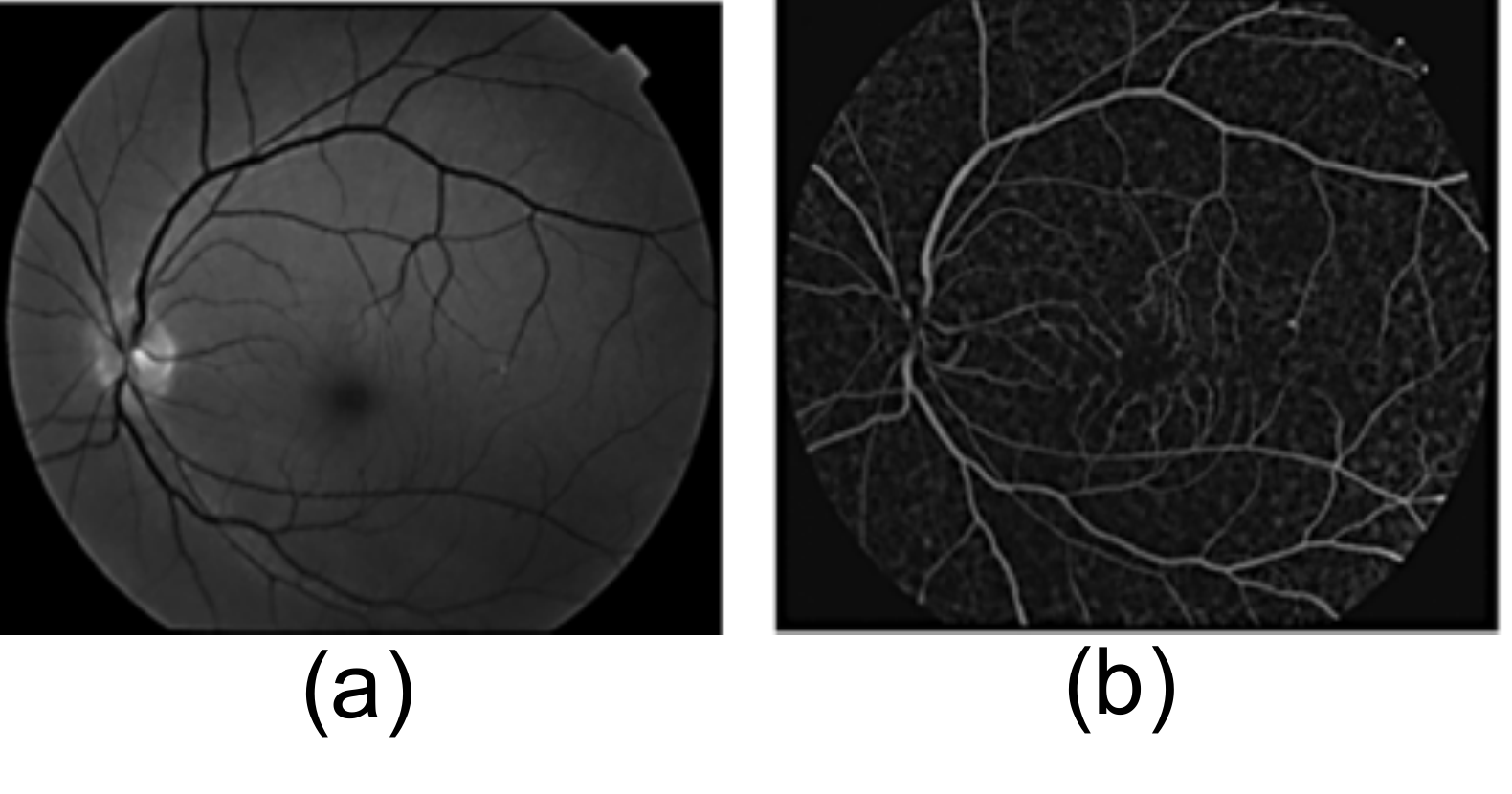}
   \end{tabular}
   \end{center}
   \caption[example] 
   { \label{fig:preprocessing} 
The result of Preprocessing step. (a) Green channel image from MUMS-DB (b) Top-hat result.}
   \end{figure}

The preprocessing step provides us with an image with high possible vessel, MAs and background contrast and also unifies the histogram of the images~\cite{pourreza2014computationally, tavakoli2017automated-fovea}. Although retinal images have three components (R, G, B), their green channel has the best contrast between vessel and background; so the green channel is selected as input image. The top-hat transform is one of the important morphological operators. In our preprocessing the basic idea is increasing the contrast between the vessels and background regions of the image. A top-hat transformation was based on a disk structure element whose diameter was empirically found that the best compromise between the features and background. The disk diameter depended on the input image resolution~\cite{tavakoli2017comparing, tavakoli2010early}. After top-hat transformation, we used contrast stretching to change the contrast or brightness of an image. The result was a linear mapping of a subset of pixel values to the entire range of grays, from the black to the white, producing an image with much higher contrast. The result of first step is shown in Fig.~\ref{fig:preprocessing}

\subsection{Vessel Segmentation and Masking} 
\begin{figure}
   \begin{center}
   \begin{tabular}{c}
   \includegraphics[height=10cm]{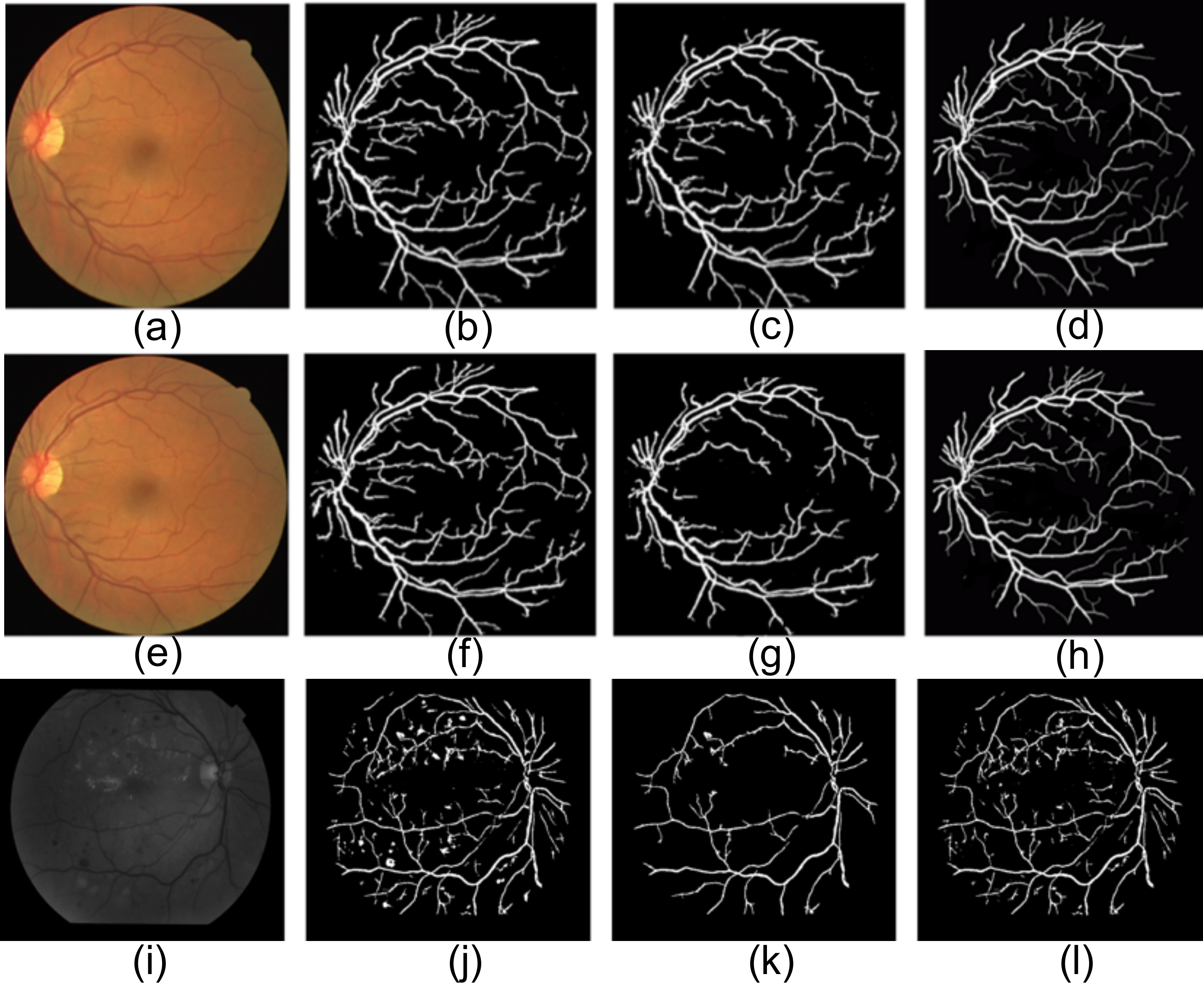}
   \end{tabular}
   \end{center}
   \caption[example] 
   { \label{fig:VesselSegmentation} 
The results for vessel segmentation in all databases using Match filter, Canny, and LoG from left to right respectively. (a-d) Normal image from DRIVE database, (e-h) Image with DR from DRIVE database, (i-l) Image from MUMS-DB.}
   \end{figure}

Here, we used three different approaches for vessel segmentation~\cite{tavakoli2017automated-onh}. The results of segmentation have been shown in Fig.~\ref{fig:VesselSegmentation}. First, we applied LoG edge detector to segment the retinal vessels. The LoG edge detector uses the second-order spatial differentiation~\cite{huertas1986detection}.
The Laplacian is usually combined with smoothing as a precursor to finding edges via zero-crossings. The 2-D Gaussian function:

	\begin{equation}
	\label{eq:Gaussian}
h(x,y)=e^{\frac{-(x^2 + y^2)}{2\sigma^2}},
	\end{equation}

Where $\sigma$ is the standard deviation, blurs the image with the degree of blurring being determined by the value of  $\sigma$. If an image is pre-smoothed by a Gaussian filter, then we have the LoG operation that is defined: $\bigtriangledown^2 G_{\sigma}\ast I$ where $\bigtriangledown^2 G_{\sigma}(x,y) = \frac{1}{2\pi \sigma^4}(\frac{x^2 + y^2}{\sigma^2} - 2)e^{\frac{-(x^2 + y^2)}{2\sigma^2}}$

In Canny edge detection~\cite{canny1986computational}, we estimate the gradient magnitude, and use this estimate to determine the edge positions and directions. 

\begin{equation}
	\label{eq:canny1}
\begin{aligned}
\begin{cases}
f_x &= \frac{\partial f}{\partial x} = K_{\bigtriangledown_x}\ast\ast(G_{\sigma}\ast\ast I)= (\bigtriangledown_x G_{\sigma})\ast\ast I, where \bigtriangledown_x G_{\sigma}=\frac{-x}{2\pi\sigma^4}e^{\frac{-(x^2 + y^2)}{2\sigma^2}}\\
\\
f_y &= \frac{\partial f}{\partial y} = K_{\bigtriangledown_y}\ast\ast(G_{\sigma}\ast\ast I)= (\bigtriangledown_y G_{\sigma})\ast\ast I, where \bigtriangledown_y G_{\sigma}=\frac{-y}{2\pi\sigma^4}e^{\frac{-(x^2 + y^2)}{2\sigma^2}}
\end{cases}
\end{aligned}
\end{equation}


Canny runs in 4 separate steps: (1) Smooth image with a Gaussian: optimizes the trade-off between noise filtering and edge localization, (2) Compute the Gradient magnitude using approximations of partial derivatives, (3) Thin edges by applying non-maxima suppression to the gradient magnitude, and (4) Detect edges by double thresholding. We can compute the magnitude and orientation of the gradient for each pixel based two filtered images~\cite{tavakoli2017automated-onh, tavakoli2017effect}.

In this paper, the Matched filter response to the detection of blood vessels is increased by proposing better filter parameters~\cite{al2007improved}. The matched filter was first proposed in to detect vessels in retinal images. It makes use of the prior knowledge that the cross-section of the vessels can be approximated by a Gaussian function. Therefore, a Gaussian-shaped filter can be used to “match” the vessels for detection. The matched filter is defined as

\begin{equation}
	\label{eq:canny2}
\begin{aligned}
G(x,y) = \frac{1}{\sqrt{2\pi\sigma^2}}e^{\frac{-x^2}{2\sigma^2}} - m_0 (y\in [-y_0,y_0])
\end{aligned}
\end{equation}
Where $m_0$ is chosen to make kernel $G(x,y)$ have zero mean.

\subsection{Microaneurysm Detection} 
To eliminate all interfering effects, we masked our images with the vascular tree mask. In this section a Matching Based approach following with concept of deep learning and CNNs was a way for extraction of circular pattern not only was utilized for MAs detection also simplified the statistical analysis of the input retinal image. 
In order to extract MAs, circular patterns with diameter lower than $125 \mu m$~\cite{walter2007automatic} should be excerpted in local sub-images (windows). Therefore, the maximum size of window was chosen twice more than size of the biggest MA. 
The size of our Matching filter was selected equal to maximum diameter of the biggest MA in pixel. Here we found that 18 pixels for MUMS-DB, and 10 pixels for DRIVE database empirically. On the other hand, point noise, end point of vessels, and bifurcations are similar to MAs (false MAs). Therefore, to validate MAs, following MAs characteristics in the images were used: Intensity, Size, and Shape. Thresholding is a way to evaluate the intensity. In other words, an easy solution to the MAs validation problem is to compare the peak amplitude intensity with predefined thresholds. Size and shape of candidate were checked.
Unlike unsupervised methods which has no initial labels~\cite{tavakoli2019quantitative} and must find natural clustering patterns in the data, deep learning approach is a learning model that can be applied for classification and regression analysis. Deep learning takes in the clustered bag of features and their corresponding labels (MAs or nonMAs) and determines the predictor clusters for each class. The process is to first train the hidden layers which is done by MA and nonMA candidates. We used CNN implemented in MATLAB~\cite{vedaldi2015matconvnet}. The purpose of the CNN classification layer is simply to transform all the net activations to a series of values that can be interpreted as probabilities in the final output layer. To do this, the CNN
MATLAB toolbox is applied onto the net outputs. 

A total number of 140 color retinal images were labeled independently with sufficient quality by an expert ophthalmologists with more than 15 years experience in diagnosing DR at early stages. Match filtering produces the MA candidates. These retinal images were made into sub-images, centered on finding the MAs. 1500 sub-images
were applied which mutually agreed with the accuracy
of their clinical label. Among these, 70\% were used
for the training purpose and remaining in the testing set.
Sub-images were extracted at twice the size of the biggest
MA ($125 \mu m$). Normal sub-images (or sub-images without
MA) were taken from the same image as the abnormal
sub-images (or sub-images with MA) in regions that were
free of MA. Overall, 470 sub-images containing MA and
the remaining sub-images without MA were taken from the
whole sub-images.

\begin{figure}
   \begin{center}
   \begin{tabular}{c}
   \includegraphics[height=8cm]{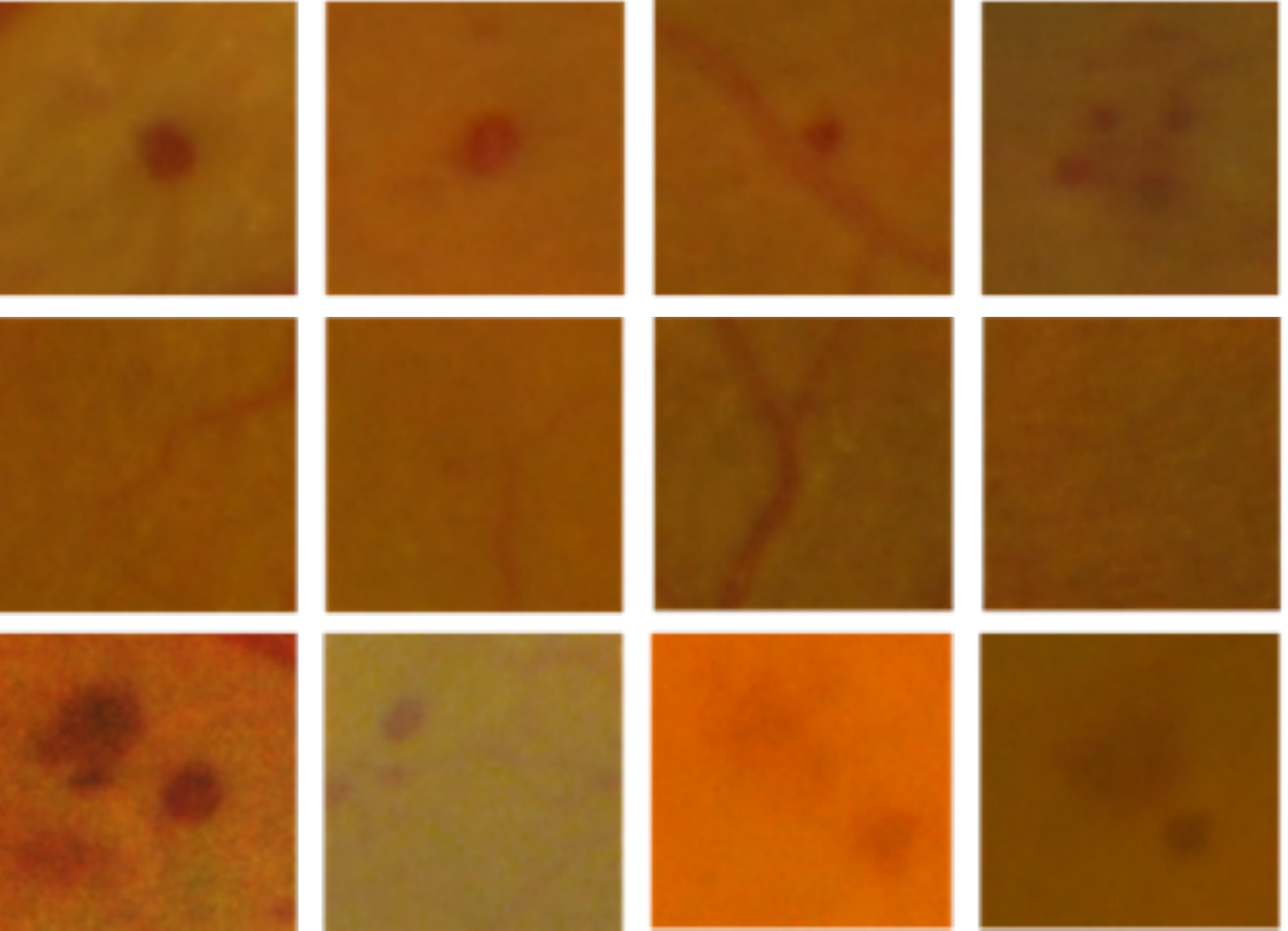}
   \end{tabular}
   \end{center}
   \caption[example] 
   { \label{fig:CNN-Candids} 
Examples of sub-images containing candidates of interest.
First row different MAs; Second row end
points of vessels and bifurcation points, Last row combination of MAs, and vessels.}
   \end{figure}

A single CNN using the MATLAB architecture with a $128 \times 128 \times 3$ input and a four-class output was designed: (1) normal, (2) MAs, (3) bifurcation points, and (4) end points of retinal vessels. Some samples from the different classes are shown in Fig. \ref{fig:CNN-Candids}. The CNN was trained on 1050 sub-images (70\%) and tested on 450 selected sub-images. Training and testing were performed using the MATLAB Deep Learning toolbox \cite{vedaldi2015matconvnet}.

\begin{figure}
   \begin{center}
   \begin{tabular}{c}
   \includegraphics[height=10cm]{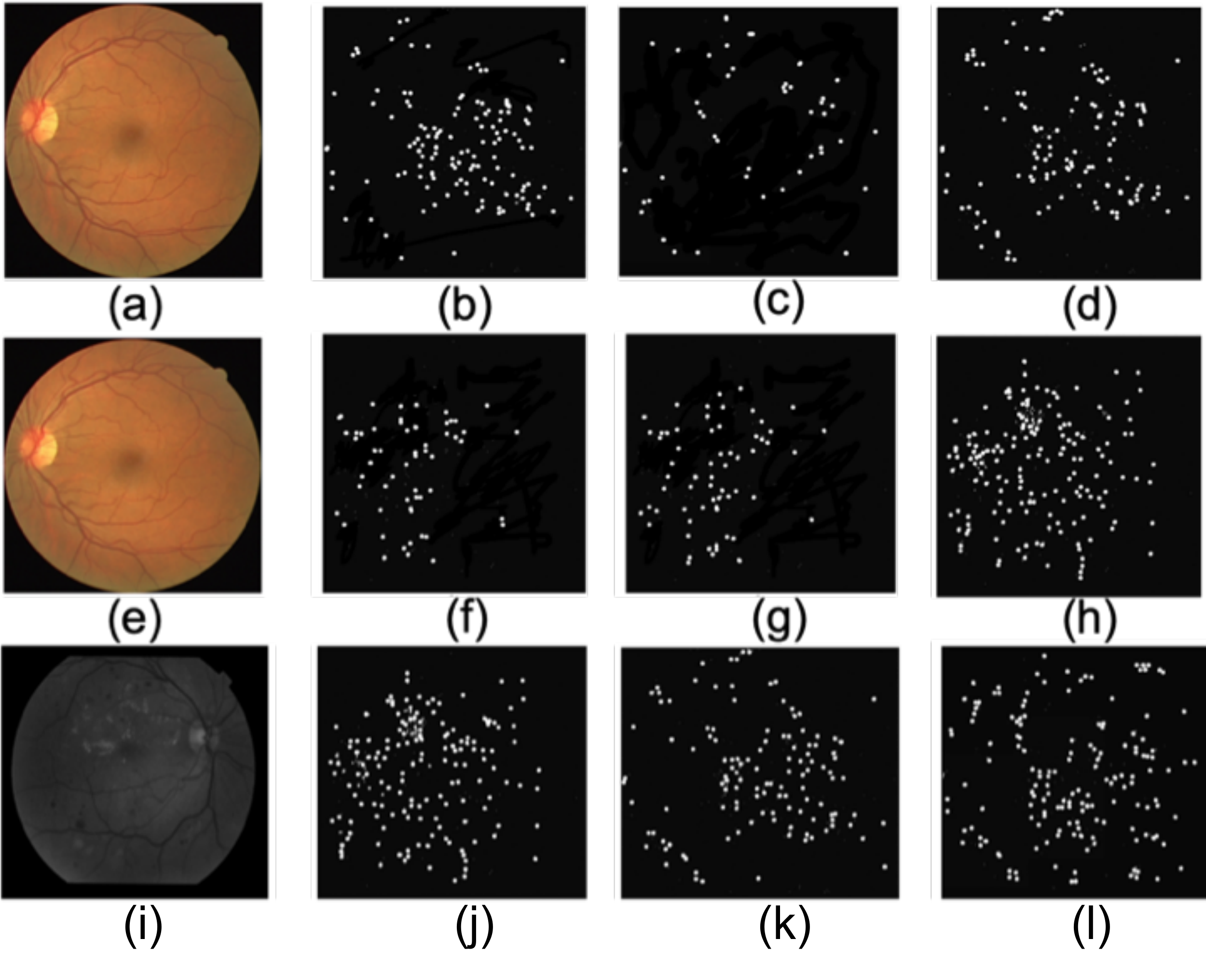}
   \end{tabular}
   \end{center}
   \caption[example] 
   { \label{fig:MA-VesselSegmentation} 
The results for MAs detection in all databases using Match filter, Canny, and LoG from left to right respectively. (a-d) image with DR from DRIVE, (e-h) Healthy image from DRIVE database, (i-l) image from MUMS-DB.}
   \end{figure}

As we mentioned, we used preprocessed images until now. Using an overlapping sliding window, the trained CNN was employed over the full scan of the image. A $20 \times 20 \times 3$ window was moved across full sub-images with a slide of 5 pixels overlapping. Each window was the input for a forward pass through our trained CNN, and produce a probability score within that sub-image for each of the four classes of normal and MAs. The result of this sliding window was a blanket of probability values over the entire image for each of the four classes. This procedure took about 1.5 minutes using a PC desktop with an Intel i5-5600HQ Processor. 
The results of MA detection has been shown in the Fig.~\ref{fig:MA-VesselSegmentation}.

\section{Results} 
In this section, statistical information about the sensitivity and specificity measures is extracted. The higher the sensitivity and specificity values, the better the procedure. 
The results for the automated method compared to the groundtruth or gold standard were calculated for each image. These metrics are defined as:

\begin{equation} \label{eq:sensitivity}
\begin{aligned}
Sensitivity  &= \frac{TP}{TP+FN} \\
Specificity  &= \frac{TN}{TN+FP} \\
\end{aligned}
\end{equation}

Where TP is true positive, TN is true negative, FP is false positive and FN is false negative same as \cite{tavakoli2013complementary, marin, tavakoli2017attenuation}.
For all retinal images, our reader labeled the MAs on the images and the result of this manual segmentation are saved to be analyzed further. According to manual MAs detection by using the LoG vessel segmentation, our automated algorithm found sensitivity of more than 85\% and specificity 80\% in detection of MAs for 100 color images in MUMS-DB and 40 images of a DRIVE database. For the Canny vessel segmentation, our automated algorithm found for both sensitivity and specificity of more than 80\% for both in detection of MAs for MUMS-DB and DRIVE databases. And lastly, using Matched filter, our algorithm found sensitivity of more than 87\% and specificity of 70\% in detection of MAs. Reaching to the sensitivity of more than 80\% makes our Computer Assisted Diagnosis (CAD) system as good as or better than other related published studies~\cite{quellec2008optimal, zhang2010detection, lazar2012retinal, habib2017detection, wu2015new}. Moreover, this sensitivity in MA based analysis show the ability of our algorithm even in treatment planning and follows up~\cite{orlando2018ensemble}. However, the disadvantage of most of the proposed approaches was that they did not avoid the overfitting issue and unable to introduce a standard feature selection principle. Bisides, for instance, in Ref~\cite{zhang2010detection} the hidden and unnoticeable structures are  ignored. Last but not least, a lot of parameters need empirically to be determined.

\section{Discussion and Conclusion} 
Since today retinal images are in digital format, it is doable to establish a computer-based system that automatically
detects landmarks and lesions from the images~\cite{tavakoli2016single}. An automatic system would save the time and workload of well-paid
clinicians letting hospitals and clinics to use their resources in other important tasks~\cite{tavakoli2019quantitative-spect}. It could also be possible
to screen more people and more often with the help of an automatic screening system, since it would be more
inexpensive than screening by humans~\cite{pourreza2009automatic, tavakoli2017automated-onh, tavakoli2017effect}.
Although the final purpose of this study was early detection of DR, we were focused only on detection of MAs. The goal of this work was to develop an automated detection of DR. So In this paper, we proposed a hierarchical method based on combination of Matching based approach and CNNs to detect all MAs from color fundus retinal image.
Altogether, for all MAs the human observer is well ahead of the automated methods. The results proved that it is possible to use our CAD system for assisting an ophthalmologist to segment fundus images into normal parts and lesions, and thus support the ophthalmologist in his or her decision making. To utilize this program in the follow up of patients, we should add an image registration algorithm so that the ophthalmologist could study the effect of his/her treatment and also the progression of the disease, not only by crisp counting, but also by spatial orientation which is included in presented method. The presented approach was evaluated through a public retinal image database DRIVE. The experiment results demonstrated that using the LoG vessel segmentation and the hierarchical approach has better detection performance in terms of of sensitivity in comparison with other published studies~\cite{quellec2008optimal, lazar2012retinal, habib2017detection, wu2015new}.




\bibliography{article}   
\bibliographystyle{spiebib}   

\end{document}